\documentclass[a4paper,11pt]{article}
\usepackage{pos}
\usepackage{graphicx} 
\usepackage{subcaption}
\usepackage{epsfig}
\usepackage{epstopdf}
\usepackage{amsfonts}
\usepackage{amsmath}
\makeatletter
\let\c@lofdepth\relax
\let\c@lotdepth\relax
\makeatother
\usepackage{tabularx}
\usepackage[T1, T2A]{fontenc}
\title{Studying the Supernova Absolute Magnitude Constancy with Baryonic Acoustic Oscillations}

\author*[a]{Denitsa Staicova}

\affiliation[a]{
$^{1}\,$ Institute for Nuclear Research and Nuclear Energy, Bulgarian Academy of Sciences, Sofia, Bulgaria}

\emailAdd{dstaicova@inrne.bas.bg}

\abstract{In this proceeding we review and expand on our recent work investigating the constancy of the absolute magnitude $M_B$ of Type Ia supernovae. In it, we used baryonic acoustic oscillations (BAO) to calibrate the supernova data and to check whether the resulting $M_B$ is constant. We used non-parametric methods like Gaussian processes and artificial neural networks to reconstruct $M_B(z)$. Here we elaborate on the results by putting them in the context of other studies investigating possible non-constant $M_B$ and the impact of the distance-duality relation.  We also present some numerical details on the calculations in the original paper and new non-parametric reconstructions, including a conservative model-independent fit, confirming its main results. Notably, we see that $M_B$ remains constant within $1\sigma$, with a possible jump around $z = 0.01 - 0.15$. Furthermore, the observed distribution of $M_B(z)$ cannot be described by a single Gaussian, displaying multiple peaks and tails. The choice of the only remaining parameter -- the sound horizon $r_d$ leads to a tension in the $M_B-r_d$ plane. Fitting different non-constant $M_B(z)$ models does not significantly improve the fit and there is no preference for any of the models by the statistical measures we employ. 
}
\FullConference{%
  Corfu Summer Institute 2023 "Tensions in Cosmology", 06-13, September, 2023\\
  Corfu, Greece}

\date{February 2023}

\begin{document}

\maketitle

\section{Introduction}
\subsection{The tensions in cosmology}
The study of supernovae type IA plays a major role in cosmology. Observations of SNe Ia have demonstrated that the universe is expanding at an accelerated rate, described by the Hubble constant, $H_0$. However, when comparing the Hubble constant measured from the early universe and from the late universe, the two measurements deviate with more than $5\sigma$ \cite{Riess:2024ohe, Abdalla:2022yfr, Vagnozzi:2023nrq} creating the so-called Hubble tension. This has led to a huge quest in improving both observational systematics and our numerical and theoretical models, but we see that there is no easy answer to the possible origin of the tension \cite{Benisty:2022psx, Dainotti:2022bzg, Dias:2023qrl, Dialektopoulos:2023jam, Alonso:2023oro, Benisty:2023vbz, Dialektopoulos:2023dhb, Briffa:2023ern, Zhai:2023yny, Bernui:2023byc, Yang:2022kho, Gariazzo:2021qtg, Bargiacchi:2023jse, Staicova:2023vxe, Dainotti:2021pqg}. Тhere are some hopeful candidates for resolving the $H_0$ tension, such as early dark energy models \cite{Poulin:2018cxd, Herold:2022iib} and interacting dark energy \cite{Pan:2019gop, DiValentino:2020kpf,DiValentino:2021izs,  Benisty:2024lmj}, but they still face challenges (for example resolving simultaneously the $H_0$ and the $\sigma_8$ tension \cite{Poulin:2022sgp, Poulin:2023lkg}). 

Different studies have been performed trying to explain the Hubble tension. Among them, some have studied the possibility of a non-constant $H_0$ across different redshifts, signaling a possible problem with the concordance model. For example, works using the redshift binned analysis on megamasers, cosmic chronometers, the Pantheon type Ia SNe and BAO  \cite{Krishnan:2020obg,Krishnan:2020vaf}, the SNe Ia Pantheon Sample (\cite{Dainotti:2021pqg}) and a non-parametric approach on SNe, BAO and CC (\cite{Jia:2022ycc}) have found evidence that $H_0$ decreases slowly with the redshift. A possible solution to the Hubble and growth tensions in the form of rapid transitions in $G_{eff}$ have been also explored \cite{Marra:2021fvf}. Intrinsic tension in the supernova sector of the local Hubble constant measurement have been discussed \cite{Wojtak:2022bct}.  Ref. \cite{Castello:2021uad} has demonstrated that a cosmological underdensity cannot solve the Hubble tension. These are just few examples of the numerous works on the topic trying to solve the problem. 
 
One way out of the situation is to look for an independent source of information such as the gravitational waves detectors. They, however, suffer for the moment from a low precision due to the small number of observed gravitational events accompanied by an electromagnetic counterpart and the associated big error box in the determination of the luminosity distance to the source and its redshift. For example, \cite{LIGOScientific:2019zcs} measured from GW170814 and GW170817 $H_0=69^{+16}_{-18}km/s/Mpc$, while \cite{Palmese:2023beh} published a reanalysis of GW170814 giving $H_0\sim 75km/s/Mpc$, \cite{Vasylyev:2020hgb} obtains from GW170817 and GW190814, $H_0=69^{+29}_{-14}$ km/s/Mpc. 
The inclusion of dark sirens (objects without EM counterpart) leads to the following results:  $H_0=76.00^{+17.64}_{-13.45}$  km/s/Mpc for 10 well-localized dark sirens and GW170817 \cite{Alfradique:2023giv}  and  $H_0=72.77^{+11.0}_{-7.55}$ for 8 dark sirens and GW170817 \cite{Palmese:2021mjm}. Clearly, the precision is still far from the necessary for shining light on the Hubble tension. Other datasets like the GRB time-delays datasets might also be used to improve our understanding, but they would be relevant only in the case of evidence for non-zero Lorentz Invariance Violation \cite{Staicova:2024ljn}.  

Back to the standard objects we use in cosmology, one tries to build  the distance ladder, starting from the closest objects (the Cepheids), through supernovae type IA (SNIA),\cite{Riess:2021jrx, Riess:2019cxk}, cosmic chronometer \cite{Moresco:2012jh,  Moresco:2015cya, Moresco:2016mzx},
the baryonic acoustic oscillations (BAO) \cite{eBOSS:2020yzd} and the cosmic microwave background \cite{Planck:2018vyg}. When including BAO in our calculations, we face a well-known problem -- the degeneracy between $H_0$, the sound horizon $r_d$ and the matter density $\Omega_m$. The sound horizon ($r_d$) is the distance at which the primordial sound waves that created BAO froze at recombination time and it's considered a known quantity. In BAO measurements, however, the sound horizon always appears coupled to $H_0$, i.e. as the common factor  $c/H_0 r_d$. This makes it impossible to find one of the quantities without calibrating the other with either the early or the late universe. There is evidence that this degeneracy spreads also to the matter density $\Omega_m$, leading to the so-called tension in the $H_0-r_d-\Omega_m$ plane \cite{Knox:2019rjx, Staicova:2021ajb, Staicova:2022zuh, Staicova:2023jic}. The coupling between the three parameters means that any solution of the $H_0$-tension should take into account also the other two parameters. 

\subsection{The supernova type IA mechanism}
Supernovae type IA (SNe Ia) are essential tool in cosmology, since they are considered standard candles for distance measurements.  These explosions originate from the thermonuclear disruption of white dwarf (WD) stars in binary systems, triggered when the mass of the WD reaches the Chandrasekhar limit of $1.4 M_{\odot}$. The limit mass gives them a stable peak luminosity and a characteristic broad, smooth light curve. The SNe Ia spectroscopy shows mostly carbon (C) and oxygen (O) lines, lack of hydrogen (H) and excess of silicon (Si).

To turn them into standard candles, one uses the distance modulus $\mu$ to connect the absolute SN magnitude $M_B$ with the measured on Earth apparent magnitude $m_B$ and the luminosity distance $d_L$ trough:
\begin{align}
&\mu(z)=m_B(z)-M_B\\
&m_B(z) - M_B=5 log_{10} [d_L(z)/1Mpc] 
\label{eq1.2}
\end{align}
Here, $z$ is the redshift of the supernova and we have omitted the color and stretching corrections and the bias term for simplicity. The luminosity distance, accounting of the expansion of the universe, $H(z)$ trough:
\begin{equation}
    d_L=(1+z) \int_0^z \frac{c dz'}{H(z')}
\end{equation}

The apparent magnitude is related to the observed flux in B-band at peak brightness $F_B$ as $$m_B \sim -2.5 log_{10}(F_B,) $$ while the observed flux depends on the emitted luminosity $L$ at distance $d_L$ as $ F=L/4\pi d_L^2$. 

The luminosity $L$ is calculated through simulations based on theoretical models of the explosion mechanism and the composition of the white dwarfs \cite{Roepke:2018gqe}. Since these models are well-established, the luminosity of a supernova is considered a known quantity, allowing their use as standard candles. However, there exist theories that may alter the Chandrasekhar limit but not the elemental composition, for example, for example magnetized WD \cite{Ablimit:2019qug}, scalar-tensor theories, exotic particles and higher dimensions \cite{Du:2022ulm, Kalita:2019yaj}. Such theories would require reevaluation of the assumptions for calculating supernova luminosity.

Furthermore, since $M_B$ is calibrated through nearby sample of objects with a known distance, it is considered a known constant. Few authors have suggested that taking a prior on $M_B$ could replace the prior on $H_0$ ~\cite{Efstathiou:2021ocp,Camarena:2021jlr}. The idea is that this would avoid double counting of low-redshift supernovae and pre-setting the deceleration parameter, $q_0$ (\cite{Riess:2016jrr}), and allowing to include the shape of the SNIa magnitude-redshift relation. 

\subsection{The constancy of $M_B$ put in a context}
The question of whether $M_B$ is a constant for SNe Ia has been explored by numerous authors. Ref. \cite{Perivolaropoulos:2022khd, Perivolaropoulos:2023iqj} developed models in which $M_B$ would show a jump at $d\sim20-50$Mpc.  Ref. \cite{Ashall:2016jxw} finds that the SNe Ia from star-forming galaxies have different mean absolute magnitude ($M_B = -19.20 \pm 0.05$ mag) than those from passive elliptical galaxies ($M_B = -18.57 \pm 0.24$ mag). Additionally, Ref. \cite{Evslin:2016koi} calibrated SN with BAO using the distance duality relation and a theoretical anchor to get $M_B(z=2.34) - M_B(z=0.32) = -0.08 \pm 0.15$.   Ref.\cite{Alestas:2020zol, Alestas:2021luu} proposed as a resolution of the Hubble tension, a model with a transition at redshift $z_t<0.1$ which performs better statistically than smooth late-time $H(z)$ deformation models.

The constancy of $M_B$ relies essentially on the distance duality relation (Etherington's reciprocity theorem ) or DDR. This relation states that the luminosity distance, $d_L(z)$, and the angular diameter distance, $d_A(z)$, are related by $d_L(z)=(1+z)^2 d_A(z)$. This relation should hold for any metric theories of gravity where the photon number is conserved and light travels along null geodesics. The validity of DDR has been studied and confirmed in numerous contexts, but some signs for its violation have come from theories with curvature or dust, also dynamical dark energy, gravitational lensing, dust extinction, modified gravity, inhomogeneities, and clustering  (\cite{Renzi:2021xii, Qin:2021jqy, Lyu2020ApJ, He:2022phb,Ma:2016bjt,Fu:2017nmw}). These studies emphasize the importance of ongoing research of the constancy of $M_B$ and its underlying assumptions.

\section{The calibration of SN with BAO }

In this proceeding, we review \cite{Benisty:2022psx} in which we tested the constancy of $M_B$ by calibrating SN measurements with BAO using a non-parametric (NP) approach. While we confirm that $M_B$ remains constant within $1\sigma$, there is an evidence of a possible jump in its value around $z = 0.01 - 0.15$ and signs of decreasing $M_B$ for high $z$. Below, we summarize our approach and our results and offer new details on it. 

\subsection{Theoretical setup}
The main idea of the approach is that having the observational data on SN and BAO, we can reconstruct in a non-parametric way the functional dependence $\mu(z)$ and $D_A/r_d(z)$ and then to use it to find $M_B$. The only remaining parameter in the problem is the sound horizon $r_d$, which we can set to either the early universe value or the late one. The formula we use to obtain $M_B$ comes from Eq. \ref{eq1.2} in which we have used the distance-duality relation to substitute the luminosity distance ($d_L$) obtained from the SN data with the angular diameter distance obtained from the BAO ($D_A$):
\begin{subequations}
\begin{equation}
    M_B = \mu_{Ia}^{} - 5\log_{10}\left[ (1+z)^2 \left(\frac{D_A^{}}{r_d^{}}\right)_\mathrm{BAO}\cdot r_d^{} \right] - 25\,.
\label{eq3a}
\end{equation}

For the theoretical uncertainty we obtain:
\begin{equation}
    \Delta M_B^{}  = \Delta\mu_{Ia}^{} + \frac{5}{\ln 10}\left[\frac{\Delta r_d^{}}{r_d^{}} + \frac{\Delta \left(D_A^{}/r_d^{}\right)_{BAO}}{\left(D_A^{}/r_d^{}\right)_{BAO}}\right]\,.
\end{equation}
\end{subequations}

\noindent where the $\Delta$ quantities refer to the observational uncertainties. This analytical formula is obtained by simplified error propagation which tends to overestimate the error (compared to the standard deviations error propagation).

In order to use this formula, we need to be able to reconstruct both $\mu_{Ia}^{}(z)$ and $\frac{D_A^{}}{r_d^{}}$. This has to be done in a model-independent way, so that that we do not introduce further tension in our results. 

\subsection{The non-parametric approach}
We consider two ways to perform a non-parametric fit of the data summarized below.

1. \textbf{The Gaussian processes}

GP reconstructs the dataset as part of a stochastic process in which each element is part of a multivariant normal distribution. It is defined via mean function $\mu(z)$ and a kernel function $k(z, z_1)$ and utilizes a Bayesian approach to optimize its kernel hyperparameters ($\sigma_f$ and $l$ ).
The GP is considered model-independent since it does not depend on the cosmological parameters, but only on the choice of the kernel. It has been tested in numerous cosmological studies, thus it is considered a robust way to make a model-independent fit. Its biggest advantage is that it naturally includes the errors of the measurements in the resulting fit. 

As kernels we consider the Radial Basis (RB) function kernel and the Rational Quadratic (RQ) kernel:
\begin{equation}\label{eq:cov-squ}
    k(z,\tilde{z})^{RB} = \sigma_f^2\exp\left(-\frac{(z-\tilde{z})^2}{2l^2}\right)\,, \\
    k(z,\tilde{z})^{RQ}  = \frac{\sigma_f^2}{\left(1 + {|z-z'}|^2/2\alpha l^2 \right)^\alpha}.
\end{equation}

Additionally, here we show data for a third kernel -- the Mattern kernel:
\begin{equation}
    k(x_i, x_j) = \frac{1}{\Gamma(\nu)2^{\nu - 1}}  \left(\frac{\sqrt{2 \nu}}{l} |z - z'|\right)^\nu K_{\nu} \left(\frac{\sqrt{2 \nu}}{l} |z - z'|\right)
\end{equation}
where $\Gamma(\nu)$ is the Gamma function of $\nu$, $|z - z'|$ is the Euclidean distance between $z$ and $z'$ and $K_(\nu)$ is the Modified Bessel function of the second kind with order $\nu$.
\begin{figure*}
 	\centering
    \includegraphics[width=0.49\textwidth]{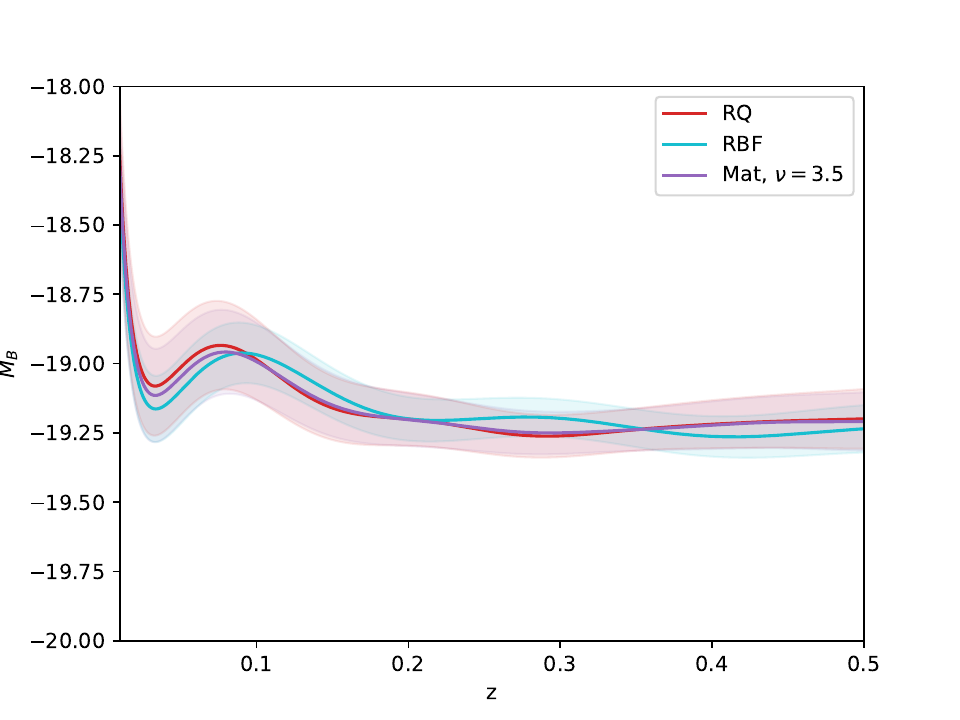}
    \includegraphics[width=0.49\textwidth]{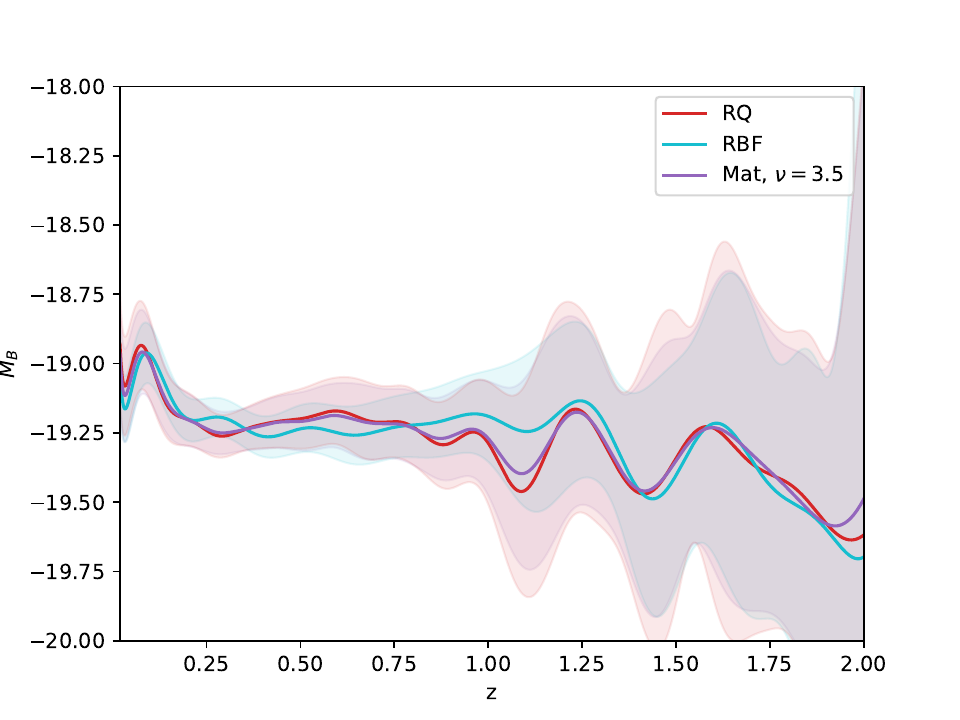}
    \caption{\it{The reconstructed value of $M_B$ for three of the GP kernels we used}} \label{fig:1}
\end{figure*}

Examples of such GP reconstruction can be seen in Fig. \ref{fig:1}. On it, we present different kernels with the same sound horizon $r_d=153$Mpc. Adding other values of the sound horizon merely shift up and down the final result, since it enters as a constant, so we do not show this here. We can see that the interval on $z$ can be split into 3 parts. At $z=0$ we have a numeric singularity due to $D_A(z=0)=0$. More importantly, we see that for small $z\sim 0.01-0.15$ we observe a jump in the values of $M_B$. It is unclear whether this is due to numerical issues or it is a feature of the data. At large $z$ we have big fluctuations due to GP becoming less certain because the BAO data points are fewer at these $z$. For this reason, we cut out plots at $z>2$. However, there is a hint of a decrease in the value of $M_B$ for larger $z$. Finally, in the middle of the reconstruction interval ($z=0.2-0.9$), we see that $M_B\sim const$ with no significant features in this interval.

2. \textbf{Artificial Neural Networks (ANNs)}

Another method to obtain a model-independent fit of the data is provided by ANNs. The ANN consists of input and output layers of neurons with a number of hidden layers of neurons, each connection having its weight. In our study, the input data are the redshifts and the output data are the observed quantities ($\mu(z),D_A/r_d(z)$ and their uncertainties. ANNs iteratively pass information forward and backward through their layers, adjusting their weights to minimize a loss function. During forward propagation, neurons in each layer compute weighted sums of inputs from connected neurons in the previous layer, followed by the application of an activation function. During the backpropagation step, the network computes gradients by propagating errors backward from the output layer to the input layer. These gradients are then used to re-adjust the weights to improve the fit. By optimizing the network's structure and parameters, ANNs provide a fit to the data without relying on specific model assumptions.

To train the ANN, we need to use mock datasets to find the optimal number of neurons and layers. In \cite{Benisty:2022psx}, the mock dataset mimics the redshift distribution of the real dataset and is generated by a risk-like statistic \cite{Wasserman:2001ng}. Furthermore, in this work we used the L1 loss function and a gradient--based optimizer based on Adam's algorithm \cite{2014arXiv1412.6980K} and the final ANN configuration is detailed in the paper.

 \begin{figure*}
 	\centering
    \includegraphics[width=0.49\textwidth]{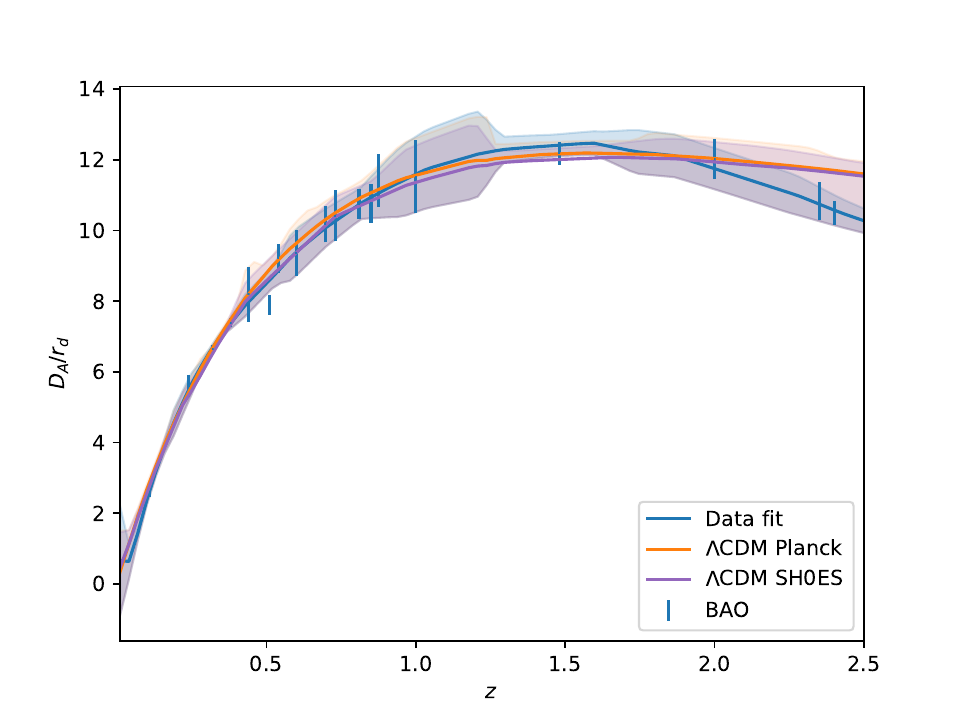}
    \includegraphics[width=0.49\textwidth]{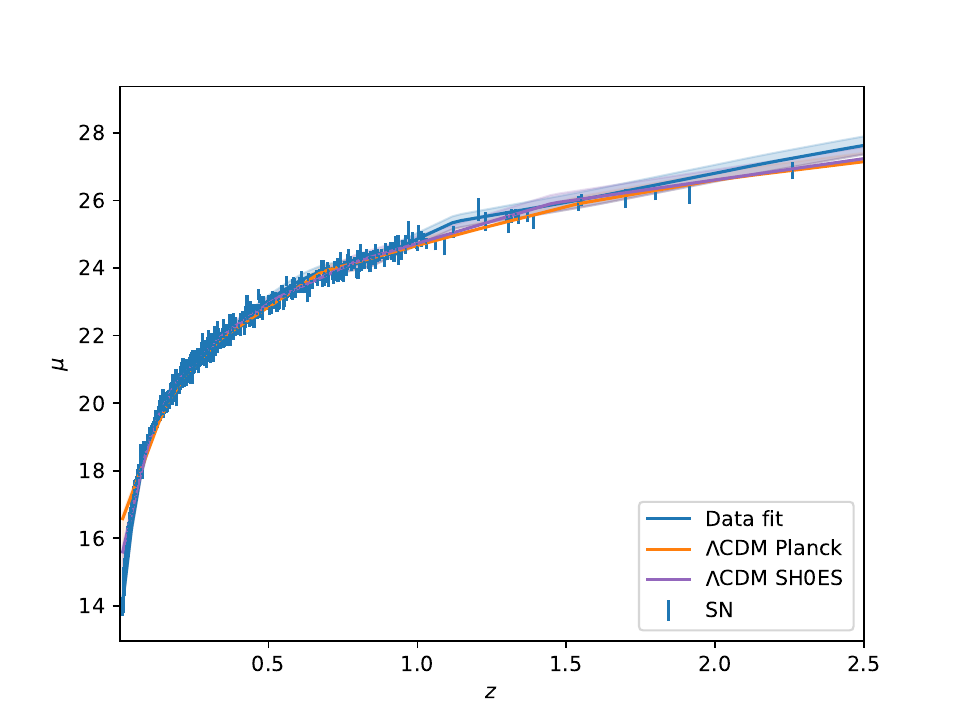}
    \caption{\it{The ANN reconstruction of $D_A(z)/r_d$ and $\mu(z)$ plotted along with the observational data with its errors. The errors correspond to $1\sigma$.}} \label{fig:3}
\end{figure*}

 \begin{figure*}
 	\centering
      \includegraphics[width=0.49\textwidth]{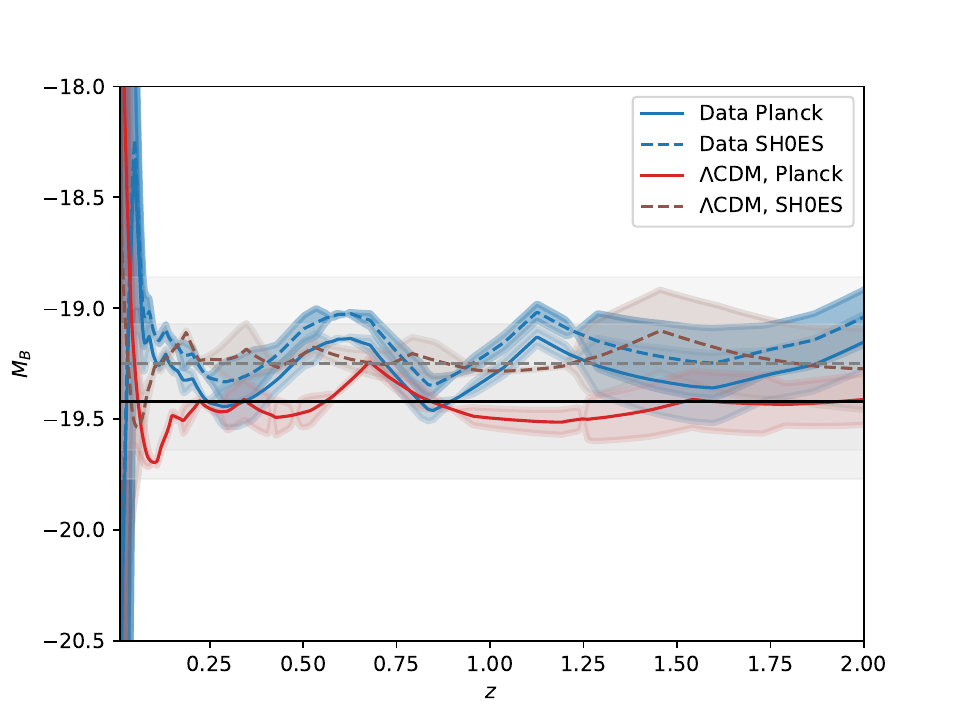}
     \includegraphics[width=0.49\textwidth]{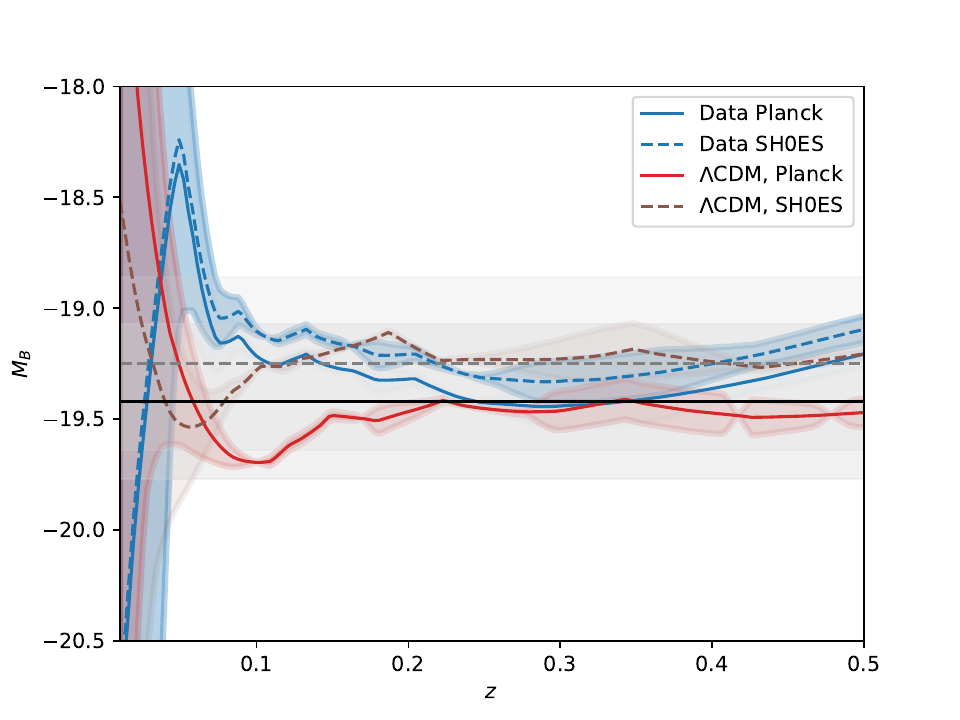}
    \caption{\it{The ANN reconstruction of $M_B(z)$ for the two $\Lambda$CDM fits and the two data fits. The two horizontal lines correspond to the expected value for $M_B$ for the SH0ES prior on $H_0$ (dashed) and the Planck one (solid)}. To the left is the full interval, to the right -- the zoom-in for small $z$} \label{fig:4}
\end{figure*}

In Fig. \ref{fig:3} we show an example of reconstructions of $D_A(z)/r_d$ and $\mu(z)$ obtained in two cases --  for mock data obtained from $\Lambda$CDM and from model-independent data fit (which we call "conservative approach"). In order to generate the mock data from $\Lambda$CDM, one needs to make an assumption for the cosmological parameters $[H_0, \Omega_m, r_d]$. Here in the Planck case we use $[67.4,0.315,147]$ \cite{Planck:2018vyg}, and for $SH0ES$: $[73.6,0.26,140.]$ \cite{Brout:2022vxf}, then one needs to add noise to the data. In the model-independent fit, we directly augment the data in a maximally conservative way (preserving the redshift distribution and the expected errors) and add noise. 

From the figure, one can see that $\Lambda$CDM differs from the conservative fit and that the BAO data is much more sensitive to changes in the model than the SN data. In  Fig. \ref{fig:4} we show the final reconstructions for $M_B$, similar to the ones obtained in \cite{Benisty:2022psx}. In it we have presented four reconstructions, the two $\Lambda$CDM-trained reconstructions that depend not-trivially on the sound horizon and the two trained on the conservative data-fit for which $r_d$ is a mere constant (since it enters only in the final formula for $M_B$). The results are very similar to the GP ones, but much cleaner with respect to the over-all error. A small difference is that the ANN jump for small $z$ is in the other direction. A notable result is that for $\Lambda$CDM-generated data, we have a significant area where the $M_B$ is constant, while for the conservative approach, $M_B$ is much more variable (since it follows closer the observational data). For all of the reconstructions, $M_B$ increases for big $z$ with the one from the conservative approach rising first. This differs from the obtained in \cite{Benisty:2022psx} probably due to the different treatment of the training data. We have also zoomed at the beginning to demonstrate the jump seen for $z\in (0,0.1]$. This jump is present in all ANN reconstructions (as it is in the GP one) and it is in an interval where we have a lot of SN data points but very few BAO data points. 

\begin{figure*}[!h]
 	\centering
    \includegraphics[width=0.49\textwidth]{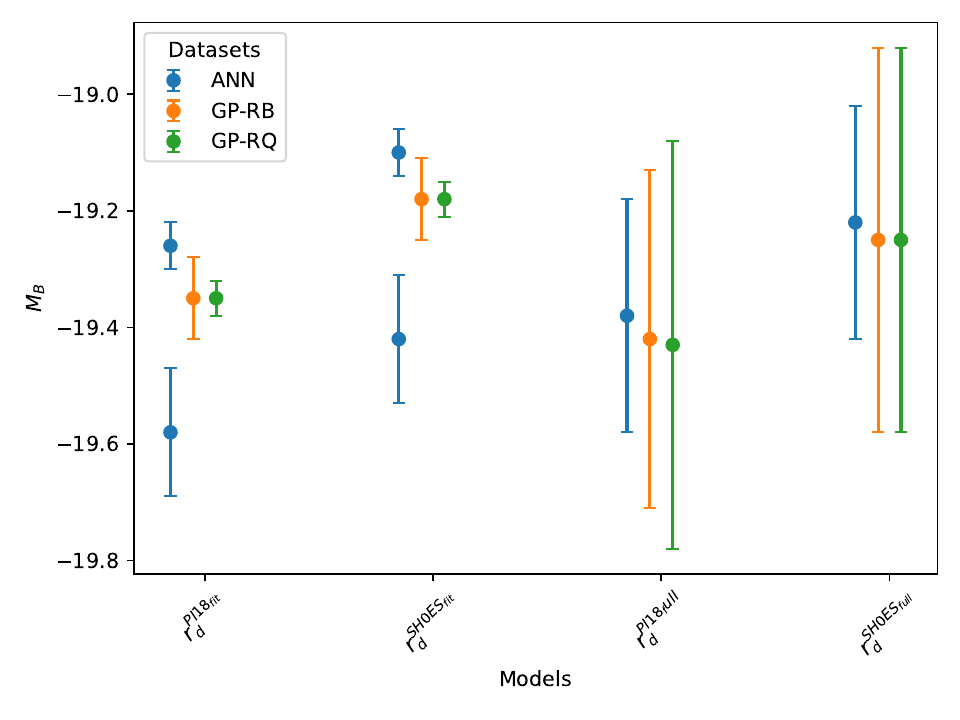}
    \caption{\it{The comparison between the final mean values and their errors for  the different methods}} \label{fig:5}
\end{figure*}

Finally, in Fig. \ref{fig:5} we have summarized the results from both NP-reconstructions published in \cite{Benisty:2022psx}. While the absolute magnitude remains constant within 1 $\sigma$, one can see that the reconstructed results do not follow a single Gaussian distribution but instead, they have two peaks, or sometimes -- tails. This have been shown on the figure by the mean values and the 1 $\sigma$ errors for the obtained results. In this plot the index "fit" refers to the results being fitted to 2 Gaussians, while "Full" refers to the results being taken as a single Gaussian. We can see that the mean values are close to the expected from Planck and SH0ES only if one assumes a single Gaussian, otherwise - they are not.  This can be seen also in Fig. \ref{fig:4} where even in the "constant" region, the reconstructions don't follow the "correct" horizontal line. 
Ref. \cite{Camarena:2019moy} using the cosmographic approach got $M_B= - 19.23\pm 0.4$. Similar results have been published recently in \cite{Mukherjee:2024akt} where the Pantheon+ dataset has been use to reconstruct $M_B$ and again a shift in the nearby $z$ region has been observed.

\subsection{Testing possible theoretical models for $M_B(z)$}

To test different possible functional dependencies of $M_B(z)$, we use the nested sampler provided by  {\it{Polychord}} \cite{Handley:2015fda} on some known forms of a nuisance parameter $\delta M_B(z)$. The goal is to check if any of these dependencies provides a better fit to the observational data, by minimizing  $\chi^2=\chi^2_{BAO}+\chi^2_{SN}$.

The functional forms for  $M_B \to M_B + \delta M_B(z)$ we consider are \cite{Tutusaus:2017ibk, Linden:2009vh}:
\begin{equation}
    \delta M_B(z) =
    \begin{cases}
        \epsilon z &  \text{Model Line}   \\
        \epsilon\left[(1+z)^\delta-1\right] & \text{Model A }  \\
        \epsilon z^\delta & \text{Model B }  \\
        \epsilon\left[\ln(1+z)\right]^\delta & \text{Model C }\,.
    \end{cases}
\end{equation}

We compare them with the default $M_B=const$ model on the background of  the flat $\Lambda$CDM model.

\begin{figure*}
 	\centering
    \includegraphics[width=0.49\textwidth]{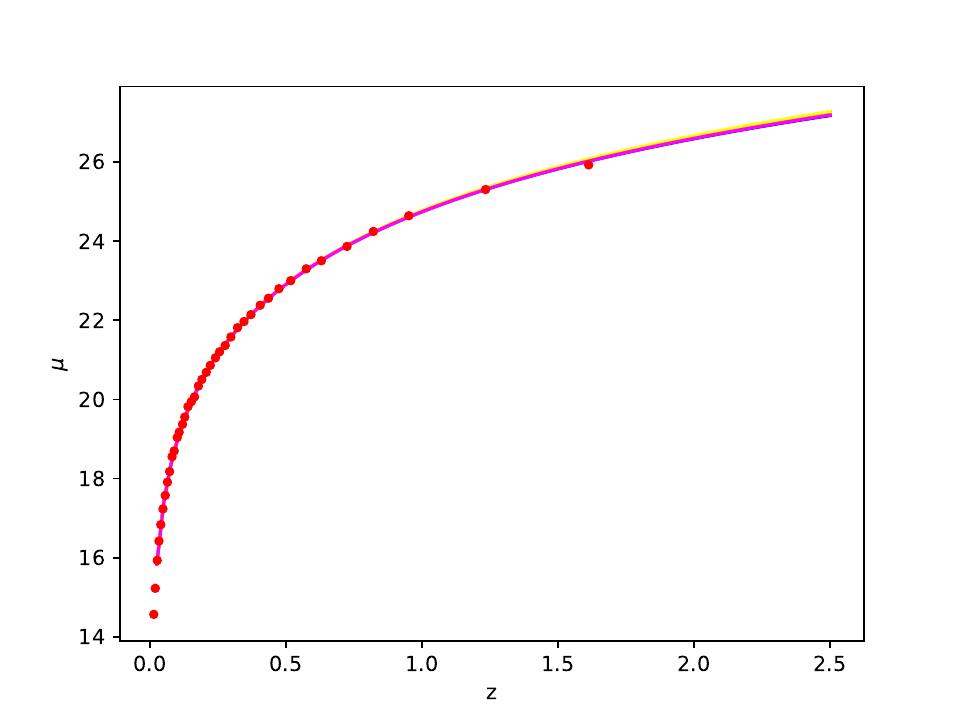}
    \includegraphics[width=0.49\textwidth]{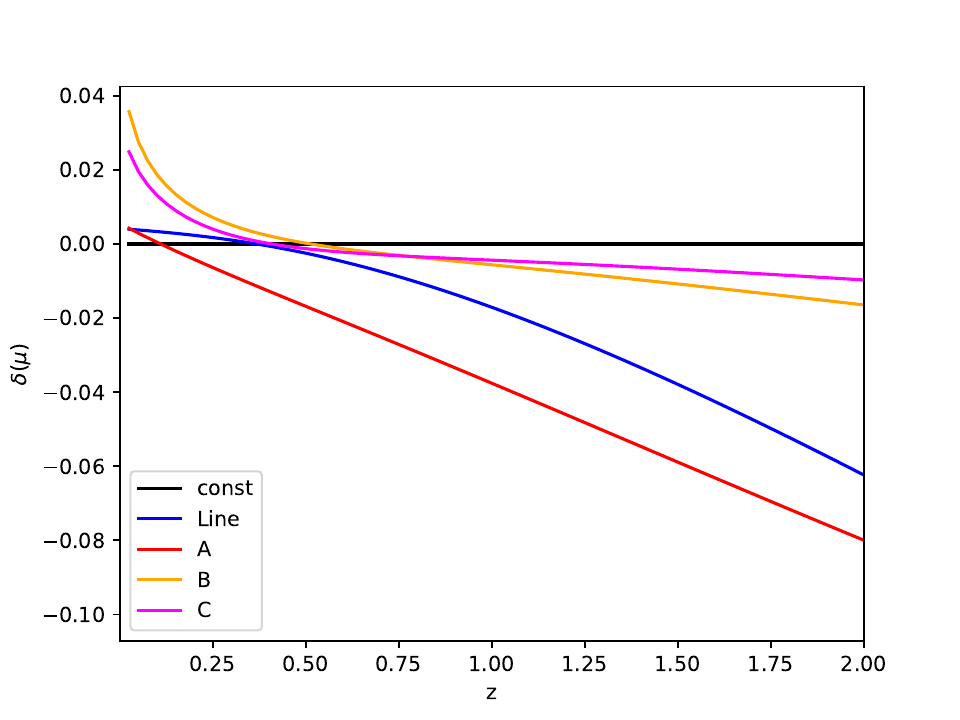}
    \caption{\it{The posterior distribution for the nuisance parameters for different models with a $\Lambda$CDM background are shown, on the right panel corresponding to a prior $r^{Pl18}_d = 147.09 \pm 0.26$ Mp, on the left one - to $r_d^{HW+SN+BAO+SH0ES} =136.1 \pm  2.7$ Mpc.}} \label{fig:6}
\end{figure*}

Our test of  possible color ($C$) and time stretching of the light curve ($X_1$) corrections in the SNIa light curves \cite{Tripp:1997wt}:
\begin{equation}
\mu_{obs} = m_B^* - (M_B - \alpha X_1 + \beta C),
\label{trippeq}
\end{equation}  show that including the color and light curve stretching corrections as constants or as Gaussian priors, gives very similar results. Here $\alpha$ is the amplitude of the stretch correction and  $\beta$ - the amplitude of the color correction. 

The priors we use to run the sampling are: uniform priors on $H_0$: $\in (50,100)$, 
$\Omega_m \in (0.2,0.4)$, $\Omega_r\in (0.,0.01)$, $\epsilon^{line}\in (-1,1)$ and $\epsilon^{A,B,C}\in (-1,1)$, $\delta\in (0,1)$.  The priors we use for $\alpha, X_1,\beta, C$ correspond to the G10 model in Ref. \cite{Pan-STARRS1:2017jku}.

A summary of the results is that none of the tested model offers a convincing fit of the data. The statistical measures we tested and particularly the Bayesian factor do not confirm the dominance of one model. As a general behavior we find that unless we fix $r_d$, the system moves too freely in the $H_0-r_d$ plane. On the other hand, fixing $r_d$ is equivalent to fixing $H_0$ and moves the final results for $M_B$ accordingly.  Adding more parameters to the model increases the error but not significantly. In Fig. \ref{fig:6}, we see that the fits from some of examined models are practically indistinguishable within the uncertainties of the $\Lambda$CDM constraints. 

\section{Conclusions and open questions}
We have reviewed our recent work \cite{Benisty:2022psx} in which we investigate the constancy of the absolute magnitude $M_B$ of Type Ia supernovae by calibrating the SN data with BAO data. In order to obtain the dependency of the angular distance on the redshift corresponding to BAO data, we have used non-parametric methods like Gaussian processes and artificial neural networks (ANN). We have shown the comparison between Gaussian processes reconstructions and ANN ones, where for the ANN we have used model-dependent and model-independent ways to generate the training data. 

The main result is that $M_B$ remains constant within $1\sigma$ CL, but there is a deviation between the theoretically expected value for $M_B$ and the reconstructed one. An interesting feature of our reconstructions is the possible jump around $z = 0.01 - 0.15$ that is seen in both NP methods we used. There is also possible evidence for an unexplained drift seen on high $z$ that may be due to the higher error in this interval of $z$ and the distribution of the BAO data. Furthermore, the observed distribution of $M_B(z)$ cannot be described by a single Gaussian, indicating multiple peaks and tails. Despite excluding $H_0$ from our analysis, one still needs to fix  $r_d$, which effectively replaces the $H_0-r_d$ tension with a tension in the $M_B-r_d$ plane. Fitting different non-constant $M_B(z)$ models does not significantly improve the fit, but also there is no preference for any of the models by the statistical measures we employ. The results of this work fit into the wider context of works examining the possibility of a non-constant $M_B$ or a drift in $H_0$. It could also have a possible impact on the study of the distance-duality relation and its possible violation. Due to the particularities of the used non-parametric approaches, however, such study needs much more observational datapoints to be trustworthy in answering questions on the cosmological tensions. 

\acknowledgments
D.S. is thankful to Bulgarian National Science Fund for support via research grant KP-06-N58/5.

\bibliographystyle{JHEP}
%
\bibliography{ref1}

\end{document}